\documentclass[12pt]{article}

\usepackage{amsmath}
\usepackage{amsfonts}
\usepackage{amssymb}
\usepackage{graphicx}
\newcommand{\be}{\begin{equation}}
\newcommand{\ee}{\end{equation}}
\newcommand{\ba}{\begin{eqnarray}}
\newcommand{\ea}{\end{eqnarray}}

\begin{document}
\begin{center}
{\bf\large
Mapping of Two-Dimensional Schr\"odinger Equation under the Point Transformation}\\
\vspace{1cm}
{\large M. V. Iof\/fe$^{1,}$\footnote{E-mail: m.ioffe@spbu.ru, corresponding author},
D. N. Nishnianidze}$^{2,}$\footnote{E-mail: cutaisi@yahoo.com},
V. V. Vereshagin$^{1,}$\footnote{E-mail: vvv@av2467.spb.edu}.\\
\vspace{0.5cm}
$^1$ Saint Petersburg State University, 7/9 Universitetskaya nab., St.Petersburg, 199034 Russia.\\
$^2$ Akaki Tsereteli State University, 4600 Kutaisi, Georgia.\\

\end{center}
\vspace{0.5cm}
\hspace*{0.5in}
\vspace{1cm}
\hspace*{0.5in}
\begin{minipage}{5.0in}
{\small For the two-dimensional Schr\"odinger equation, the general form of the point transformations such that the result can be
interpreted as a Schr\"odinger equation with effective (i.e. position dependent) mass is studied. A wide class of such models with different
forms of mass function is obtained in this way. Starting from the solvable two-dimensional model, the variety of solvable partner
models with effective mass can be built. Several illustrating examples not amenable to conventional separation of variables are given.}

\end{minipage}

\section{Introduction}

The non-relativistic quantum models with mass which depends on coordinates (effective mass models, or models with position-dependent mass) appeared in many different branches of modern physics such as nuclear physics \cite{nucl}, quantum wires and dots \cite{-25}, quantum liquids \cite{-22}, physics of semiconductors \cite{-21-1}-\cite{-21-4}, etc.
These models were investigated by different approaches such as direct solution of Schr\"odinger equation, separation of variables in multidimensional case, some algebraic methods and others (see, for example, \cite{kocak} -
\cite{quesne-2015}). In particular, the novel popular method \cite{one-plastino}
- \cite{quesne-2016} was provided by supersymmetrical intertwining relations which are the main
ingredient of Supersymmetrical Quantum Mechanics methods \cite{susy-1} - \cite{susy-4}. Most of these studies concerned the simplest class of problems - one-dimensional ones \cite{susy-pt-1}
- \cite{susy-pt-2}. However, keeping in mind possible physical applications, it is desirable to extend the class of solvable models onto cases of higher dimensionality of space. Some results in this direction were already obtained in the literature \cite{tkachuk-1}, \cite{quesne-1}, \cite{mustafa}, \cite{quesne-2}, \cite{CIN}, \cite{schulze}, \cite{nikitin}, \cite{quesne-2016}, \cite{quesne-2015}, \cite{IKN}, but, as a rule, the systems with separation of variables were considered.

In the present paper, the method of point transformations is applied for construction of two-dimensional solvable models with position-dependent mass (PDM) starting from the initial Schr\"odinger equation with constant mass. This method is not new: it was used for investigation of Schr\"odinger equations by many authors. Mainly, they deal either with one-dimensional problems \cite{susy-pt-1},  \cite{susy-pt-3}, \cite{one-milanovic}, \cite{tkachuk-2}, \cite{kocak}, \cite{one-gonul}, \cite{one-ganguly}, \cite{susy-pt-2}, or with multidimensional ones \cite{tkachuk-1}, \cite{quesne-1}, \cite{mustafa}, \cite{quesne-2}, \cite{CIN}, \cite{nikitin}, \cite{IKN} which allow the separation of variables from the very beginning (as a rule, for the spherical symmetry of equation). Here, we will consider two-dimensional PDM problems which generally speaking are not amenable to separation of variables. This approach will allow to build a wide class of corresponding solvable Schr\"odinger equations with PDM and modified form of potential. The spectra of obtained problems coincide with those of the initial ones (with constant mass), and the expressions for the wave functions can be derived explicitly. The paper is organized as follows. The basic formulas of the two-dimensional point transformations method are given in Section 2. Section 3 contains results for the different forms how the mass function depends on coordinates. Several specific illustrative examples are also given in this Section. Some final comments are included in Conclusions.

\section{Formulation of the method}

Let us start from the conventional two-dimensional Schr\"odinger equation with potential $V(\vec x)$ and constant mass (it will be taken equal $1/2$ for simplicity):
\begin{equation}\label{1}
\bigl(-\partial_i^2 + V(\vec x)\bigr)\Psi(\vec x)=E\Psi(\vec x);\,\, \vec x=(x_1,x_2); \,\, \partial_i\equiv \partial/\partial x_i,
\end{equation}
(here and below summation over repeated indices $i=1,2$ is implied). We perform the point transformation of coordinates $x_i$ with simultaneous change of wave function:
\begin{equation}\label{pt}
y_i\equiv y_i(\vec x);\quad \vec y=(y_1,y_2); \quad \Psi(\vec x)\equiv g(\vec y)\widetilde \Psi(\vec y),
\end{equation}
where at the moment, $y_i$ and $g$ are arbitrary functions of their arguments. The idea is to use this transformation in order to obtain the partner PDM Schr\"odinger equation of the following form:
\be
\bigl(-\partial_{y_i}\frac{1}{M(\vec y)}\partial_{y_i} + U(\vec y)\bigr)\widetilde\Psi(\vec y)=E\widetilde\Psi(\vec y),   \label{ef}
\ee

After the direct substitution of (\ref{pt}) into (\ref{1}) we impose two conditions: 1) the second order kinetic term has the form of Laplacian with positive scalar multiplier $1/M(\vec y)$, and 2) the coefficients in front of the first derivatives over $y_i$ in the l.h.s. of (\ref{1}) must have the same form as those contained in the kinetic term of (\ref{ef}).
The first condition includes three equations:
\begin{equation}\label{delta}
(\partial_iy_k)(\partial_iy_n) =\frac{1}{M(\vec y)}\delta_{kn}.
\end{equation}
Two first equations in (\ref{delta}) - with $k=n$ - can be rewritten as:
\ba
&&\partial_1y_1= \frac{1}{M^{1/2}(\vec y)}\cos\alpha(\vec y); \,\, \partial_2y_1= \frac{1}{M^{1/2}(\vec y)}\sin\alpha(\vec y); \label{alpha} \\
&&\partial_1y_2= \frac{1}{M^{1/2}(\vec y)}\cos\beta(\vec y); \,\, \partial_2y_2= \frac{1}{M^{1/2}(\vec y)}\sin\beta(\vec y), \label{beta}
\ea
and third one - for $k\neq n$ - provides the relation $\alpha = \beta \pm \pi/2.$ Below the upper sign will be used for simplicity, but the final result for the lower sign is the same. Eqs.(\ref{alpha}), (\ref{beta}) lead to:
\be
\partial_1y_1=-\partial_2y_2;\quad \partial_2y_1=\partial_1y_2, \label{ab}
\ee
and therefore,
\be
\partial_i\partial_iy_1=\partial_i\partial_iy_2=0. \label{bc}
\ee
Summing up, equations (\ref{delta}) are solved by means of the arbitrary harmonic functions:
\be
y_1=f(z)+ f^*(z^*);\quad y_2=\widetilde f(z)+\widetilde f^*(z^*), \label{f}
\ee
where $z\equiv x_1+ix_2,\, z^*\equiv x_1-ix_2,$ the star sign corresponds to the complex conjugation. In terms of $z,\,z^{\star},$ Eq.(\ref{delta}) for $k=n$ provides
the condition for the functions $f,\,\widetilde f:$
\be
f'(z)f^{\star\prime}(z^*)=\tilde f'(z)\widetilde f^{\star\prime}(z^*)=\frac{1}{4M}.
\nonumber
\ee
Therefore,
\be
\frac{f'(z)}{\widetilde f'(z)}=\frac{\widetilde f^{\star\prime}(z^*)}{f^{\star\prime}(z^*)}\equiv \omega ;\quad  \omega = const,
\nonumber
\ee
and substituting (\ref{f}) into (\ref{delta}) with $k\neq n$, one obtains that $\omega$ is pure imaginary number $\omega =\pm i,$ and $\widetilde f = \pm if + \gamma,$ with $\gamma$ - an arbitrary real constant which can be eliminated by means of imaginary constant shift of $f(z).$  Thus, the solution of Eqs.(\ref{delta}) has the form:
\ba
&&y_1=f(z)+f^*(z^*);\quad y_2=\pm i(f(z)-f^*(z^*)); \label{yy}\\
&&\frac{1}{M(\vec y)}=4f'(z)f^{\star\prime}(z^*) \label{11}
\ea
(for simplicity, we shall consider the upper sign in $y_2$). Depending on the specific form of $f(z),$ variables $y_{1,2}$ may vary either over the whole plane $\vec y,$ or in its part only. Both opportunities are realized in examples of next Section.

The second condition for the point transformation (\ref{pt}) concerns the first derivatives in the new Schr\"odinger equation (\ref{ef}).
This restriction means that:
\be
-\frac{2}{M(\vec y)} \frac{\partial_{y_k}g(\vec y)}{g(\vec y)}=-\partial_{y_k}(\frac{1}{M(\vec y)}),  \label{eq}
\ee
i.e. up to a constant multiplier:
\be
g^2(\vec y)=1/M(\vec y)=4f'(z)f^{\star\prime}(z^*), \label{gM}
\ee
and the potential $U$ in (\ref{ef}) is:
\be
U(\vec y)=V(\vec x(\vec y))-\frac{\partial_{y_i}\partial_{y_i}g(\vec y)}{g(\vec y)}
=V(\vec x(\vec y))-\frac{f''(z)f^{\star\prime\prime}(z^*)}{4(f'(z)f^{\star\prime}(z^*))^2}, \label{U}
\ee
where in the last term, the variables $z,\,z^*$ has to be expressed in terms of $\vec y=(y_1, y_2)$ by means of relations (\ref{yy}). This can be done for a wide class of functions $f(z).$

The behaviour of wave functions $\widetilde\Psi(\vec y)$ at the boundary depends on the function $g(\vec y),$ and it is regulated by relation (\ref{pt}). To establish the correspondence between quantum problems Eq.(\ref{1}) with potential $V(\vec x)$ and Eq.(\ref{ef}) with potential (\ref{U}), we have to check the conditions of normalizability of their wave functions. The normalizability of the wave functions $\widetilde\Psi(\vec y)$ (see the definition (\ref{pt})) means:
\be
\int \mid\widetilde\Psi(\vec y)\mid^2 d^2y =
\int M(\vec y)\mid\Psi(\vec x(\vec y))\mid^2 d^2y < \infty ,\label{norm}
\ee
and the condition of Hermiticity of PDM Hamiltonian is expressed \cite{quesne-1} through the asymptotic behaviour at the boundary of coordinate space:
\be
\frac{\mid\widetilde\Psi(y_1,y_2)\mid^2}{\sqrt{M(y_1,y_2)}} \to 0.    \label{herm}
\ee
The condition (\ref{norm}) can be transformed as:
\ba
&&\int M(\vec y)\mid\Psi(\vec x(\vec y))\mid^2 d^2y
=\int M \mid\Psi(\vec x)\mid^2 det\mid \frac{\partial y_i}{\partial x_k} \mid d^2x = \nonumber\\
&&=\int M \mid\Psi(\vec x)\mid^2 (4f'(z)f^{\star\prime}(z^*)) d^2x
= \int \mid\Psi(\vec x))\mid^2 d^2x. \label{trans}
\ea
Therefore, by the point transformations derived above, the normalized wave functions $\Psi(\vec x)$ are transformed to the normalized wave functions $\widetilde\Psi(\vec y)$
of the corresponding model with PDM. The condition (\ref{herm}) has to be checked separately for specific forms of mass function $M.$

\section{Examples}

Everything was prepared above for the description of variety of quantum models with PDM which are obtained from the conventional Schr\"odinger equation by means of the point transformations. Practically, the only possible difficulty is an opportunity to express the initial variables $x_1,\, x_2$ in terms of new variables $y_1,\, y_2$ for the chosen function $f(x_1+ix_2)$ (see (\ref{yy})). Below, we shall present several examples with different forms of $\vec y-$dependence of the mass function $M(y_1,y_2).$ It is convenient to express both the mass function $M(\vec y),$ and the second term in the effective potential (\ref{U}) in terms of a new function $F :$
\ba
&& F(f)\equiv 1/f'(z);\quad F^*(f^*)\equiv 1/f^{\star\prime}(z^*); \label{F}\\
&&M(\vec y)=\frac{1}{4}F(f)F^*(f^*);  \,\,
U(\vec y)=V(\vec x(\vec y))-\frac{F'F^{\star\prime}}{4FF^*} \label{UU}
\ea
Below, we shall give a few examples either starting from the definite dependence of the mass function on coordinates $\vec y,$ or starting from the definite form of the function $f(z).$

{\bf I.} Let us choose the case when the mass depends only on one coordinate: $M(\vec y)=M(y_1).$ The corresponding function $F(f)$ can be found from the condition $\partial_{y_2}M(y_1,y_2)=0,$ i.e.
\be
\partial_f(F(f)F^*(f^*))=\partial_{f^*}(F(f)F^*(f^*)).
\label{I}
\ee
The solution of (\ref{I}) is $F(f)=2\gamma \exp(\alpha f(z))$, and from the definition (\ref{F}), one obtains that:
\be
f(z)=\frac{1}{\alpha}\ln(\frac{\alpha}{2\gamma}z+\delta). \label{a}
\ee
Then, the mass function has an exponential form:
\be
M(\vec y)=|\gamma|^2\exp{(\alpha y_1)} \label{Ia}
\ee
($\alpha$ - an arbitrary real constant), and the second term in the effective potential $U$ of (\ref{U}) is constant in this case:
\be
U(\vec y)=V(\vec x(\vec y))-\alpha^2/4. \label{Ib}
\ee

For illustration, let us consider the function $f(z)$ with real $\gamma$ and $\delta =0,$ when
\ba
&&x_1=\frac{2\gamma}{\alpha}\exp(\alpha y_1/2)\cos(\alpha y_2/2);\nonumber\\ &&x_2=-\frac{2\gamma}{\alpha}\exp(\alpha y_1/2)\sin(\alpha y_2/2), \nonumber
\ea
the mass function has the form (\ref{Ia}), and the variable $\vec y$ belongs to a strip along $y_1.$

Up to now, the potential $V(\vec x)$ in the initial constant mass problem (\ref{1}) was arbitrary. Here for illustrative example, let us take the simple non-isotropic harmonic oscillator potential:
\be
V(\vec x(\vec y))=\omega_1^2x_1^2+\omega_2^2x_2^2
=\frac{2\gamma^2\exp(\alpha y_1)}{\alpha^2}\biggl[(\omega_1^2-\omega_2^2)\cos(\alpha y_2)+(\omega_1^2+\omega_2^2)\biggr], \label{Ie}
\ee
and the effective potential (\ref{Ib}) is periodic along $y_2$ and exponentially growing along $y_1$ from zero to infinity. The problem with such potential and PDM (\ref{Ia}) can not be solved by means of standard separation of variables, but it can be solved analytically using the point transformation approach developed above since its partner problem (\ref{1}) is exactly solvable. According to (\ref{ef}), the wave functions are expressed in terms of Hermite polynomials $H_n$:
\ba
&&\widetilde\Psi_{n_1, n_2}(y_1,y_2)=\sqrt{\frac{1}{2^{n_1+n_2}n_1!n_2!}} \bigl(\frac{\omega_1\omega_2}{\pi^2}\bigr)^{1/4}|\gamma|\exp{(\frac{1}{2}\alpha y_1)} \nonumber\\
&&\exp{[-\frac{1}{2}(\omega_1x_1^2(\vec y)+\omega_2x_2^2(\vec y))]}H_{n_1}(\sqrt{\omega_1}x_1(\vec y))H_{n_2}(\sqrt{\omega_2}x_2(\vec y)),
\label{ex-1}
\ea
where $n_1, n_2$ are two integer non-negative numbers, and old variables $x_1, x_2$ are expressed in terms of $\vec y$ as given above. These wave functions are normalizable according to (\ref{norm}), (\ref{trans}), and the asymptotical condition (\ref{herm}) is fulfilled for them. For illustration, the lowest wave functions are presented in Figures 3-5 for particular values of parameters.

\vspace{4pt}
\begin{center}
\includegraphics[width=0.8\textwidth]{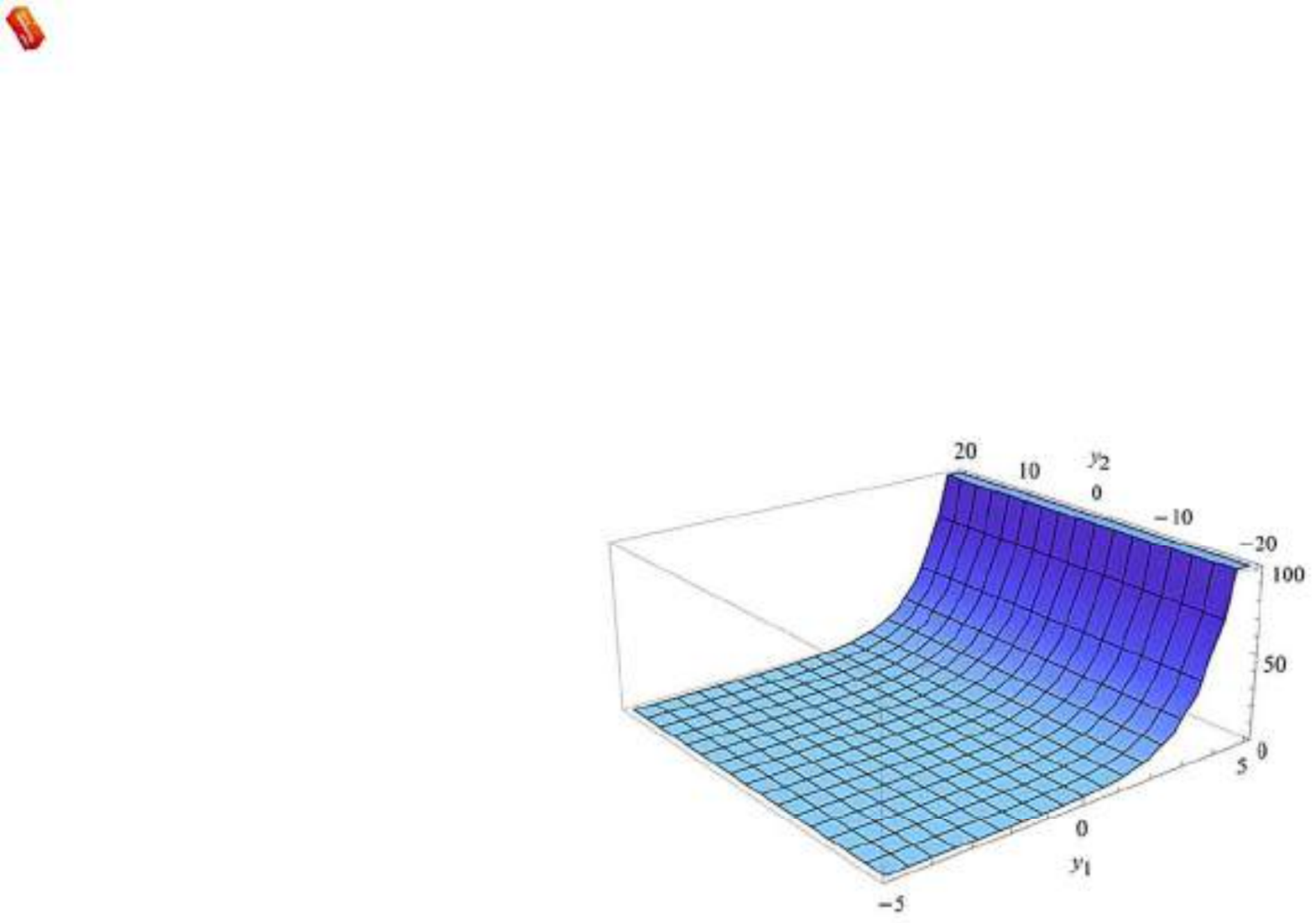}

\noindent{\it Fig.1} Mass function $M(\vec y)$ of (\ref{Ia}) for Example I with $\alpha =\gamma =1$.
\end{center}

\vspace{2pt}

\begin{center}
\includegraphics[width=0.8\textwidth]{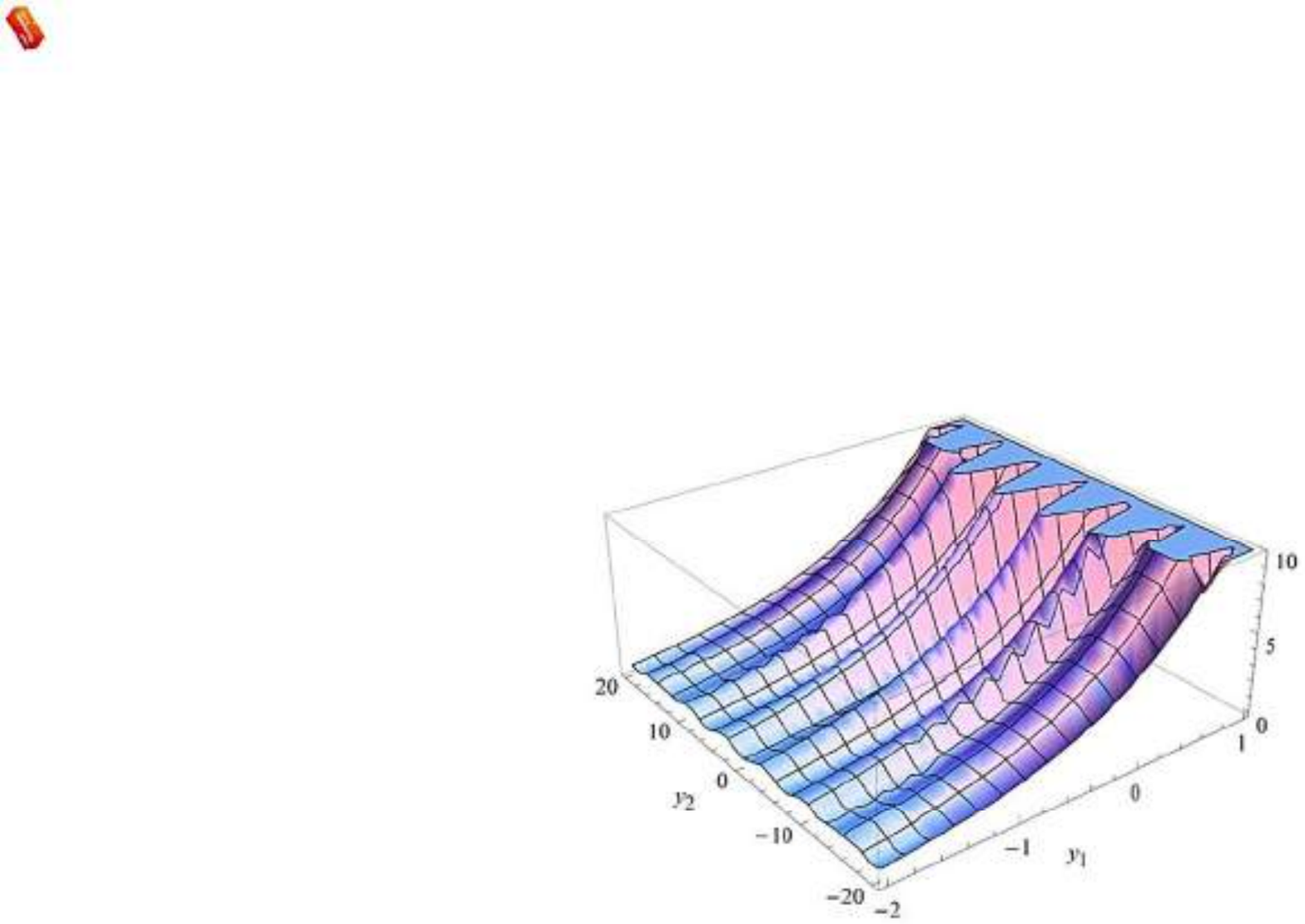}

\noindent{\it Fig.2} Potential $U(\vec y)$ of (\ref{Ib}),(\ref{Ie}) for Example I with $\alpha =\gamma =1$ and $\omega_1=1,\, \omega_2=\sqrt{2}$.

\end{center}

\vspace{2pt}

\begin{center}
\includegraphics[width=0.8\textwidth]{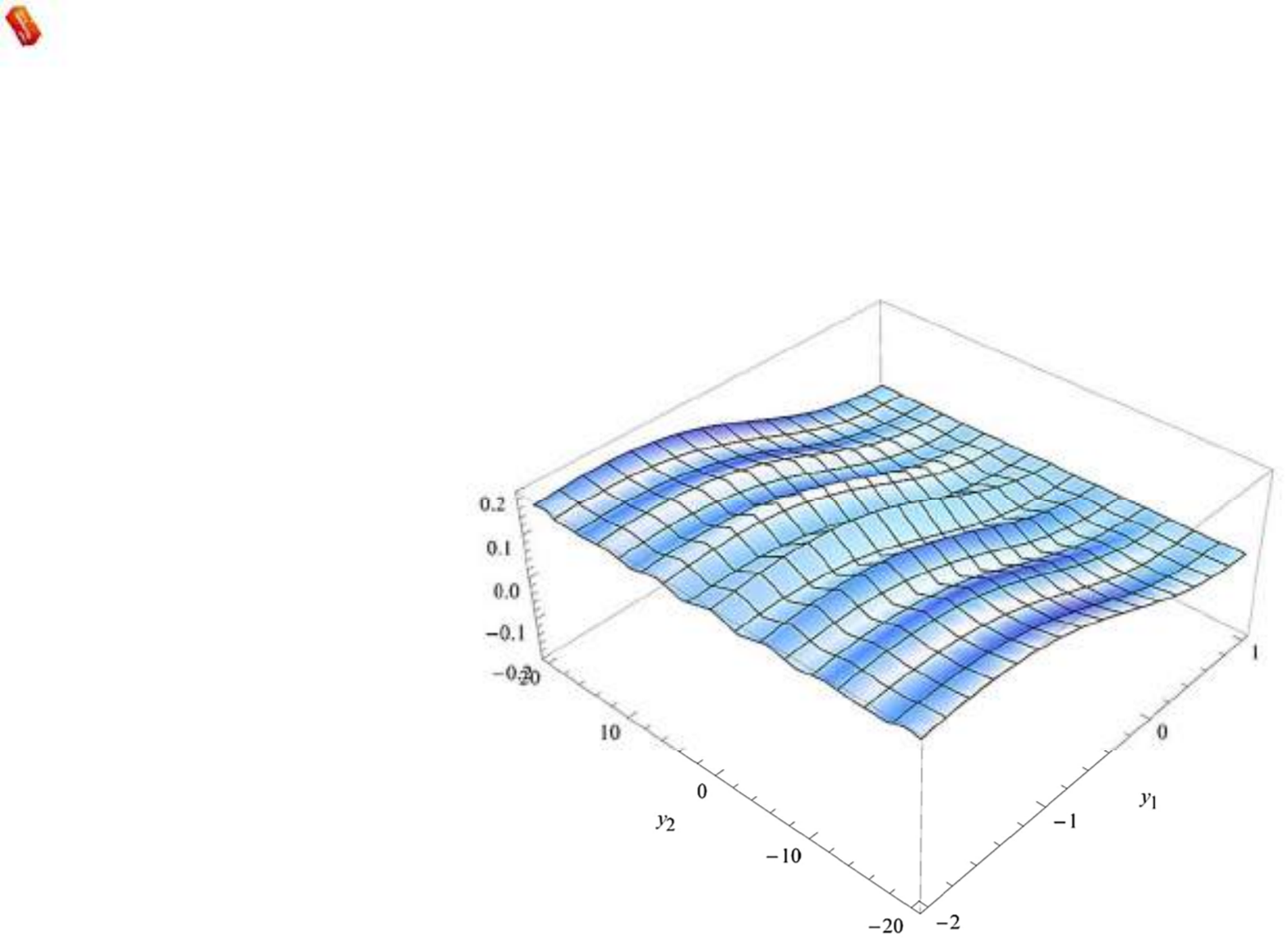}

\noindent{\it Fig.3} Ground state wave function $\widetilde\Psi_{0,0}(\vec y)$ (see (\ref{ex-1})) for Example I with $\alpha =\gamma =1$ and $\omega_1=1,\, \omega_2=\sqrt{2}$.

\end{center}

\vspace{2pt}
\begin{center}
\includegraphics[width=0.8\textwidth]{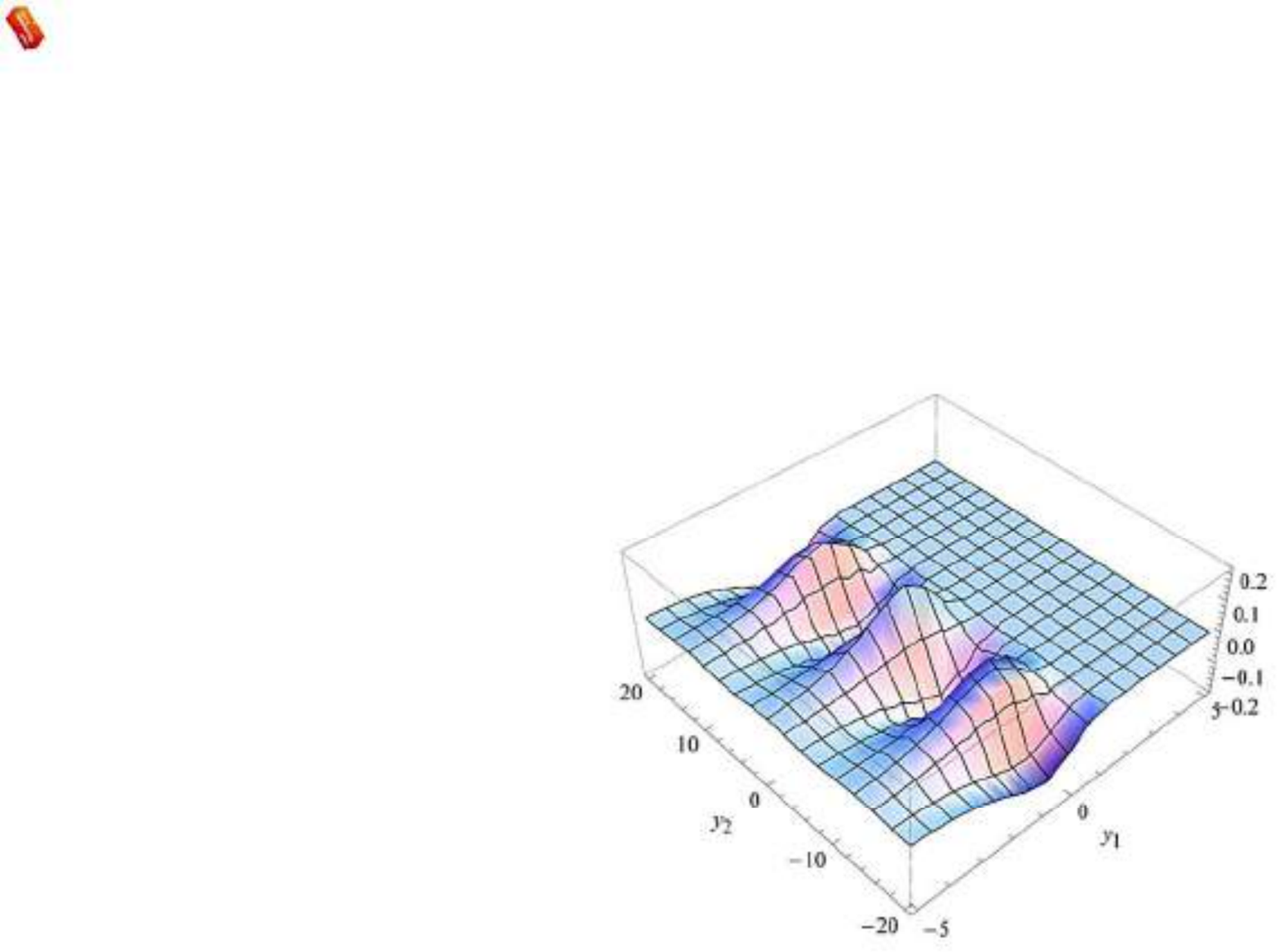}

\noindent{\it Fig.4} Wave function $\widetilde\Psi_{1,0}(\vec y)$ (see (\ref{ex-1})) of the first excited state for Example I with $\alpha =\gamma =1$ and $\omega_1=1,\, \omega_2=\sqrt{2}$.
\end{center}

\vspace{2pt}
\begin{center}
\includegraphics[width=0.8\textwidth]{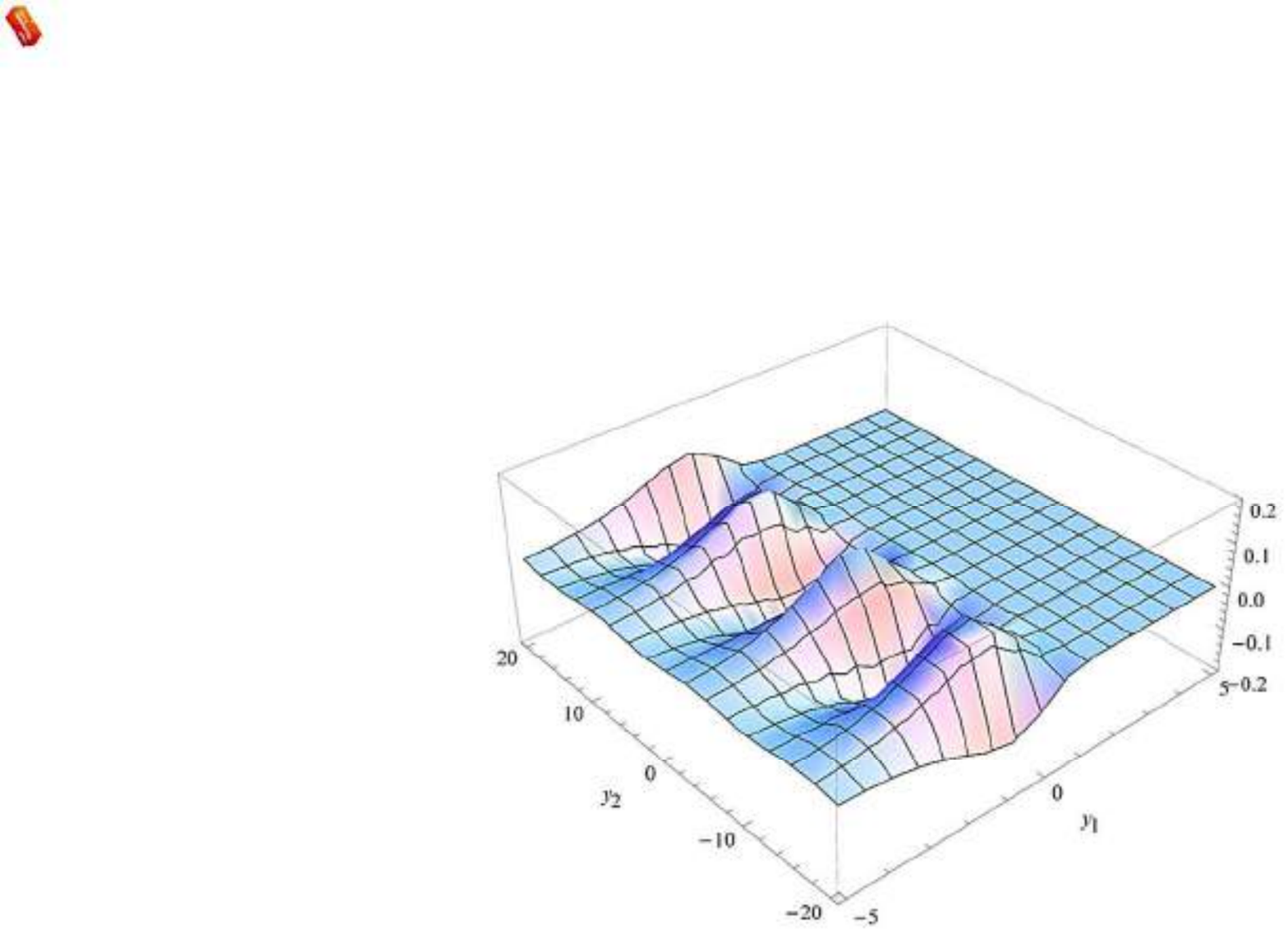}

\noindent{\it Fig.5} Wave function $\widetilde\Psi_{0,1}(\vec y)$ (see (\ref{ex-1})) of the second excited state for Example I with $\alpha =\gamma =1$ and $\omega_1=1,\, \omega_2=\sqrt{2}$.

\end{center}

\vspace{2pt}

{\bf II.} Let the mass function to be a sum of two terms: $M(y_1,y_2)=M_1(y_1)+M_2(y_2),$ i.e. $\partial_{y_1}\partial_{y_2}M(y_1,y_2)=0.$ In terms of $F(f):$
\ba
&&(\partial_f^2-\partial_{f^*}^2)(FF^*)=0;\,\, F''/F=F^{\star\prime\prime}/F^*=\lambda^2=Const;\nonumber\\
&&F(f)=2A\cosh(\lambda f+B). \label{II}
\ea
After the suitable constant shift of $y_1,\,y_2,$ the mass function becomes:
\be
M(y_1,y_2)=\frac{2|A|^2}{\lambda^2}[\cosh(\lambda y_1)+\cos(\lambda y_2)], \label{IIa}
\ee
with asymptotically decreasing behaviour along $y_1$ and periodic along $y_2.$
Also, one obtains:
\ba
&&x_1=\frac{2A}{\lambda}\sinh(\lambda y_1/2)\cos(\lambda y_2/2);\nonumber\\ &&x_2=-\frac{2A}{\lambda}\cosh(\lambda y_1/2)\sin(\lambda y_2/2). \label{IIc}
\ea
Taking again for illustration the potential of non-isotropic harmonic oscillator $V(\vec x)=\omega_1^2x_1^2+\omega_2^2x_2^2,$ one obtains the effective potential $U,$ not amenable to separation of variables:
\ba
&&U(\vec y)=\frac{4A^2}{\lambda^2}\bigl[\omega_1^2\sinh^2(\lambda y_1/2)\cos^2(\lambda y_2/2)
+\omega_2^2\cosh^2(\lambda y_1/2)\sin^2(\lambda y_2/2)\bigr] -\nonumber\\
&&-\frac{\lambda^2[\cosh(\lambda y_1)-\cos(\lambda y_2)]}{4[\cosh(\lambda y_1)+\cos(\lambda y_2)]}, \label{IIb}
\ea
but solvable by means of point transformation from the harmonic oscillator.

{\bf III.}
In our point transformation approach, $M(\vec y)=M(\rho^2)$ can be obtained as solution of the condition:
\be
(y_1\partial_{y_2}-y_2\partial_{y_1})M(\vec y)=0, \nonumber
\ee
i.e. $(f\partial_f-f^*\partial_{f^*})M=0.$ Therefore,
\be
\frac{fF'}{F}=\frac{f^*F^{\star\prime}}{F^*}=\lambda =const;\quad F(f)=\beta \nonumber f^{\lambda}.
\ee
The general solutions for $f(z)$ and, correspondingly, for $M(\vec y)$ depend on the value of real constant $\lambda :$
\ba
&&f(z)=\bigl(\frac{\lambda +1}{\beta}z+(\lambda +1)\alpha\bigr)^{1/(\lambda +1)};\nonumber\\
&&M(\vec y)=\frac{\beta^2}{4^{\lambda +1}}(\rho^2)^{\lambda};\,\, \lambda\neq -1;\label{IIIb} \\
&&f(z)=\gamma\exp(z/\beta ); \quad  M(\vec y)=\beta^2\rho^{-2}; \quad \lambda =-1.\label{IIIc}
\ea
The effective potential $U$ is:
\be
U(\vec y)=V(\vec x(\vec y))-\frac{\lambda^2}{\rho^2}. \label{IIId}
\ee

The specific illustrative models of such kind can be considered.

1). Let us take (\ref{IIIb}) with $\lambda =-2;\, \alpha =0;\, \beta =-b= real ,$ so that:
\be
f(z)=b/z;\,\, x_1=2by_1/\rho^2;\,\, x_2=2by_2/\rho^2;\\,\, M(\vec y)=4b^2/\rho^4. \label{IIIe}
\ee
Taking again the simple potential $V(\vec x)$ of non-isotropic harmonic oscillator, it must be expressed now in terms of polar coordinates $\rho,\,\varphi $
in the plane $\vec y:$
\be
V(\vec x(\vec y))=\omega_1^2x_1^2+\omega_2^2x_2^2
=\frac{2b^2}{\rho^2}[(\omega_1^2-\omega_2^2)\cos(2\varphi )+(\omega_1^2+\omega_2^2)]. \label{IIIf}
\ee
Thus, the Schr\"odinger equation with PDM from (\ref{IIIe}) takes the form:
\be
\biggl(-\frac{\rho^4}{4b^2}(\partial_{\rho}^2+\frac{5}{\rho}\partial_{\rho}+
\frac{1}{\rho^2}\partial_{\varphi}^2)
+\frac{2b^2[(\omega_1^2-\omega_2^2)\cos(2\varphi )+(\omega_1^2+\omega_2^2)]}{\rho^2}-\frac{4}{\rho^2}\biggr)\widetilde\Psi(\vec y)
=E\widetilde\Psi(\vec y). \label{IIIg}
\ee
It is clear that this equation is not amenable to separation of variables, but in the point transformations approach (\ref{IIIg}) is equivalent to the
Schr\"odinger problem (\ref{1}) with constant mass and the oscillator potential (\ref{IIIf}). This problem is obviously solvable by means of separation
of variables in terms of $x_1,x_2.$ Both the spectrum $E_n$ and wave functions $\Psi_n(\vec x)$ are well known through the Hermite polynomials \cite{landau}.
Now, the solution $\widetilde\Psi(\vec y)$ of the Schr\"odinger equation (\ref{ef}) with PDM in the form (\ref{IIIg}) can be obtained according to (\ref{pt}),
and its spectrum coincides with the well known spectrum of non-isotropic oscillator \cite{landau} (it is easy to check that the condition (\ref{herm}) is also fulfilled).
Thereby, the nontrivial problem with PDM becomes solvable due to the point transformations method described above.

2) The second specific model with the same initial oscillator potential (\ref{IIIf}), but with the choice $\lambda =-1/2;\, \alpha =0;\, \beta =\frac{1}{2a}= real$ in (\ref{IIIb}) gives:
\be
f(z)=az^2=\frac{y_1-iy_2}{2};\quad M(\vec y)=\frac{1}{8|a|\rho}. \label{IIIh}
\ee
In this case, the Schr\"odinger equation with such PDM can be written as::
\be
\biggl(-8|a|\rho \bigl(\partial_{\rho}^2+\frac{2}{\rho}\partial_{\rho}+\frac{1}{\rho^2}\partial_{\varphi}^2\bigr)+ U(\rho, \varphi)\biggr)\widetilde\Psi(\vec y)=E\widetilde\Psi(\vec y), \label{IIIi}
\ee
where the effective potential is:
\be
U(\rho, \varphi)= \frac{\rho}{4a}\biggl((\omega_1^2-\omega_2^2)\cos(\varphi)+(\omega_1^2+\omega_2^2)\biggr)-\frac{1}{4a^2\rho^2}. \label{IIIk}
\ee
Again, the Schr\"odinger problem with PDM can not be solved straightforwardly, but it is amenable to analytical solution by means of the point transformation
approach. Similarly to the Example I, the wave functions for the Schr\"odinger equation (\ref{IIIi}) are expressed in terms of Hermite polynomials:
\ba
&&\widetilde\Psi_{n_1, n_2}(y_1,y_2)= \nonumber\\
&&=\sqrt{\frac{1}{2^{n_1+n_2}n_1!n_2!}} \bigl(\frac{\omega_1\omega_2}{\pi^2}\bigr)^{1/4}\frac{1}{\sqrt{8|a|\rho}}
\exp{\biggl[-\frac{1}{2}(\omega_1x_1^2+\omega_2x_2^2)\biggr]} H_{n_1}(\sqrt{\omega_1}x_1)H_{n_2}(\sqrt{\omega_2}x_2)=\nonumber\\
&&=\sqrt{\frac{1}{2^{n_1+n_2}n_1!n_2!}} \bigl(\frac{\omega_1\omega_2}{\pi^2}\bigr)^{1/4}\frac{1}{\sqrt{8|a|\rho}}
\exp{\biggl[-2\rho\biggl(\omega_1\cos^2(\varphi/2)+\omega_2\sin^2(\varphi/2)\biggr)\biggr]}\cdot\nonumber\\
&&\cdot H_{n_1}\biggl(2\sqrt{\omega_1\rho}\cos(\varphi/2)\biggr)H_{n_2}\biggl(2\sqrt{\omega_2\rho}\sin(\varphi/2)\biggr),
\label{ex-3}
\ea
where $n_1, n_2$ are two integer non-negative numbers, and according to (\ref{IIIh}), the variables $x_1=2\sqrt{\rho}\cos{(\varphi/2)};\,x_2=2\sqrt{\rho}\sin{(\varphi/2)}$ are expressed in terms of the polar coordinates $\rho, \varphi $ on the plane $\vec y$. Using the properties of Hermite polynomials, it is not difficult to check that although a part of these solutions (when both $n_1$ and $n_2$ are even) is singular at the origin, all wave functions (\ref{ex-3}) are normalizable on the plane $\vec y.$ The asymptotical condition (\ref{herm}) is also satisfied.
For illustration, the mass function $M(\vec y)$, the singular potential (\ref{IIIk}) and three lowest wave functions $\widetilde\Psi_{0,0},\,\widetilde\Psi_{1,0}$ and $\widetilde\Psi_{0,1}$ are given in Figures 6-10 for some values of parameters.

\vspace{4pt}

\begin{center}
\includegraphics[width=0.8\textwidth]{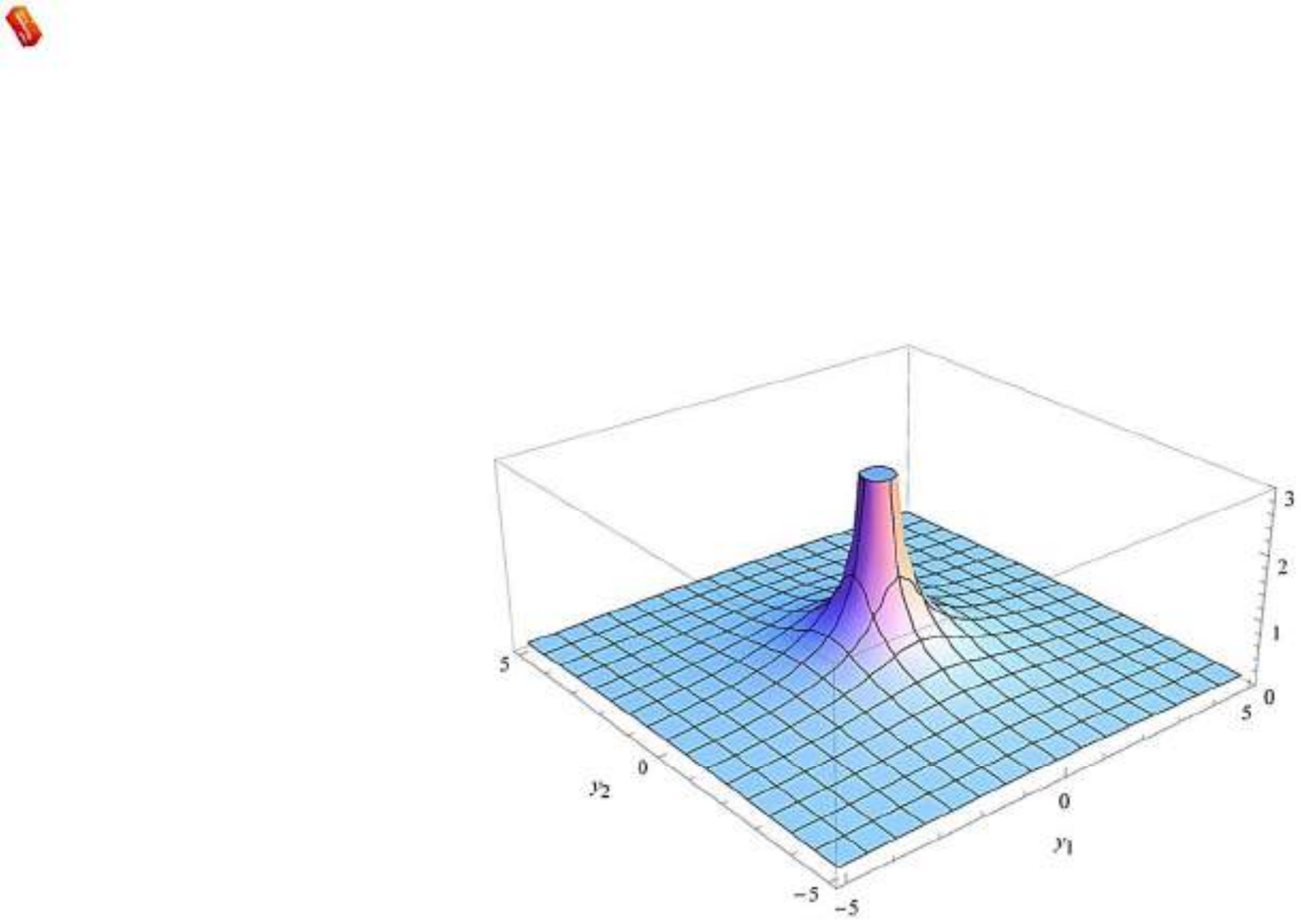}

\noindent{\it Fig.6} Mass function $M(\vec y)$ (\ref{IIIh}) for Example III (case 2) with $a=1/8$.

\end{center}

\vspace{2pt}

\begin{center}
\includegraphics[width=0.8\textwidth]{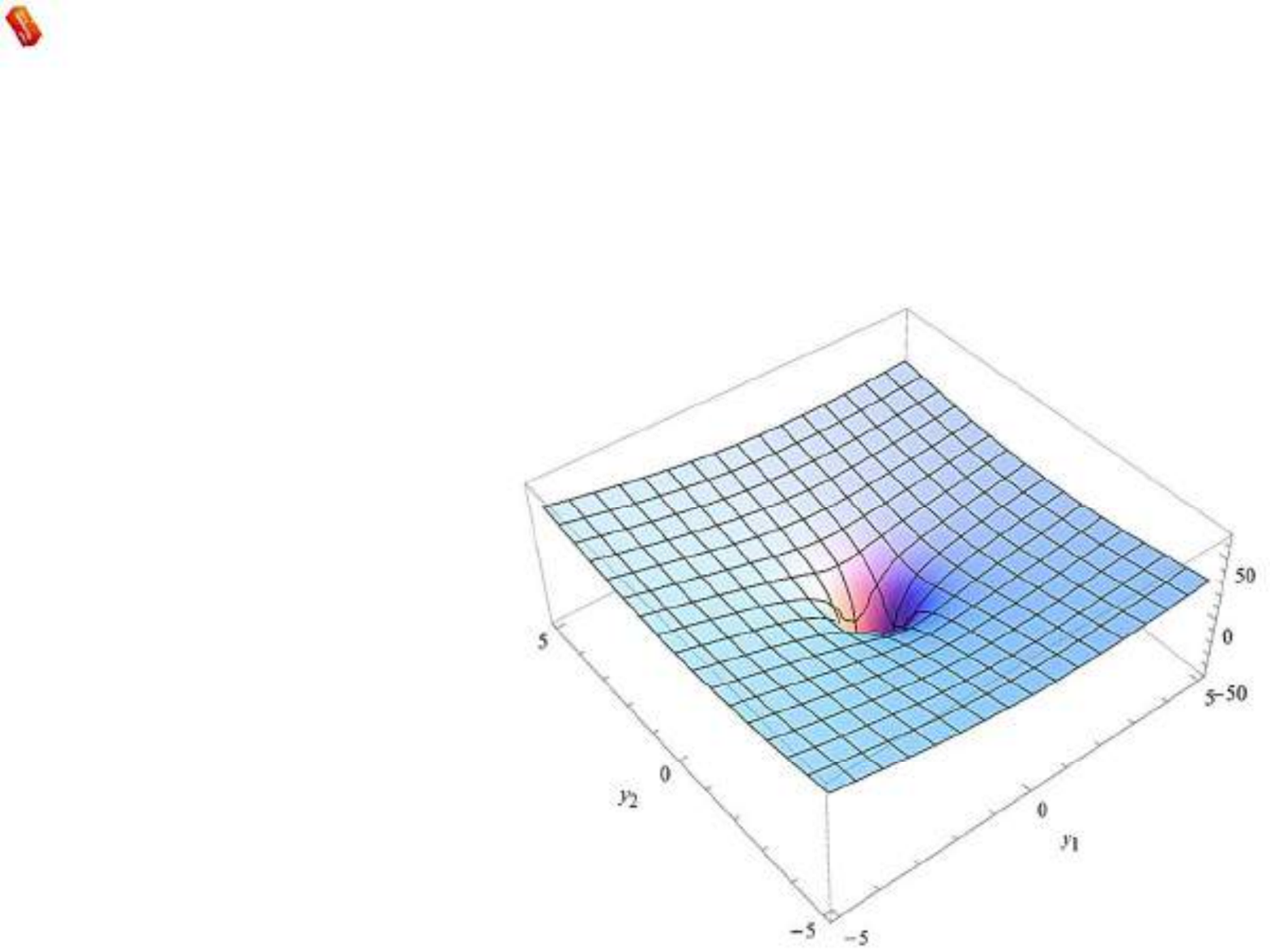}

\noindent{\it Fig.7} Potential (\ref{IIIk}) for Example III (case 2) with $a=1/8$ and $\omega_1=1,\, \omega_2=\sqrt{2}$.

\end{center}

\vspace{2pt}

\begin{center}
\includegraphics[width=0.8\textwidth]{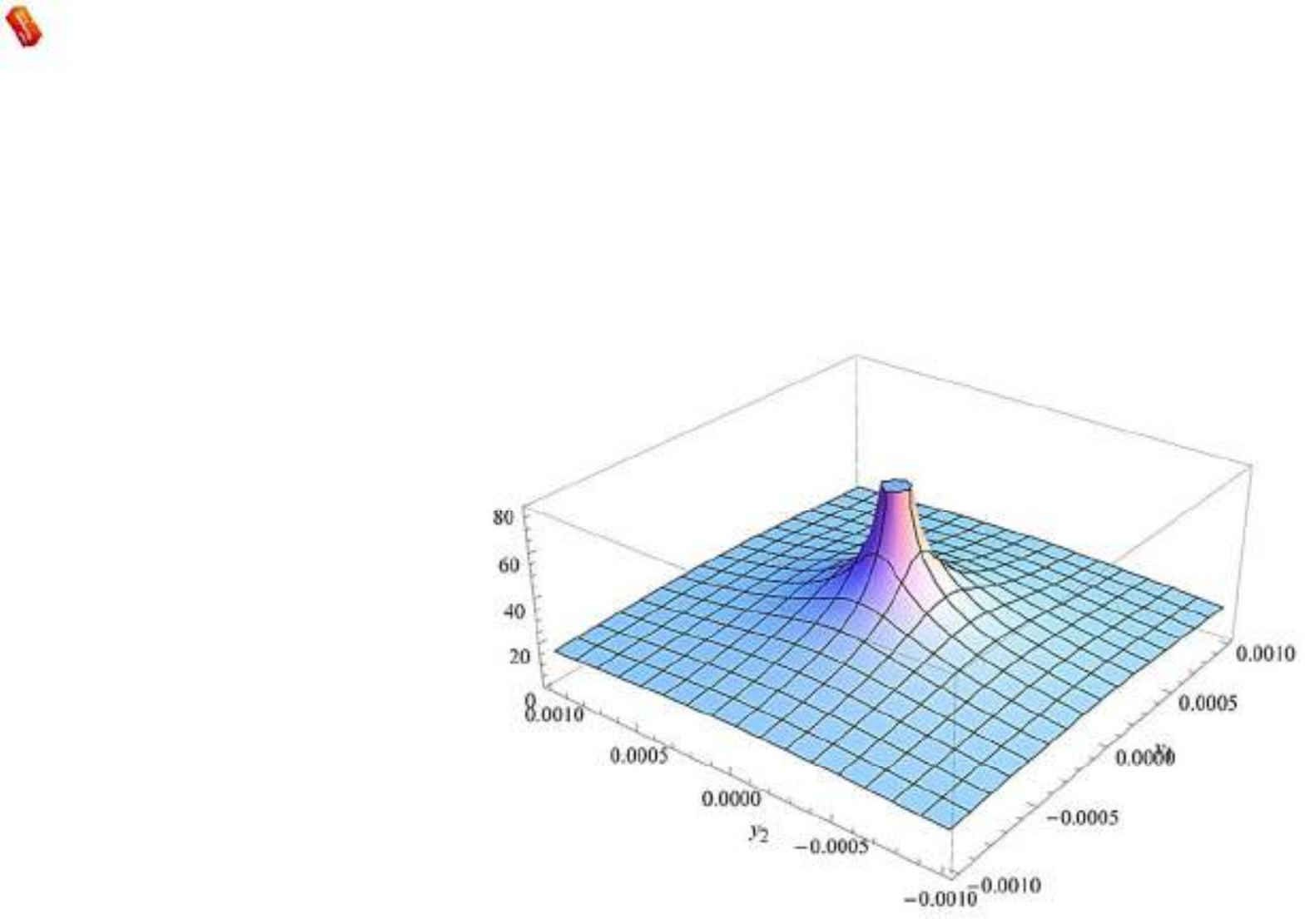}

\noindent{\it Fig.8} Ground state wave function $\widetilde\Psi_{0,0}(\vec y)$ (see (\ref{ex-3})) of the Schr\"odinger equation (\ref{IIIi}) with potential (\ref{IIIk}) for Example III (case 2) with $a=1/8$ and $\omega_1=1,\, \omega_2=\sqrt{2}$.

\end{center}

\vspace{2pt}
\begin{center}
\includegraphics[width=0.8\textwidth]{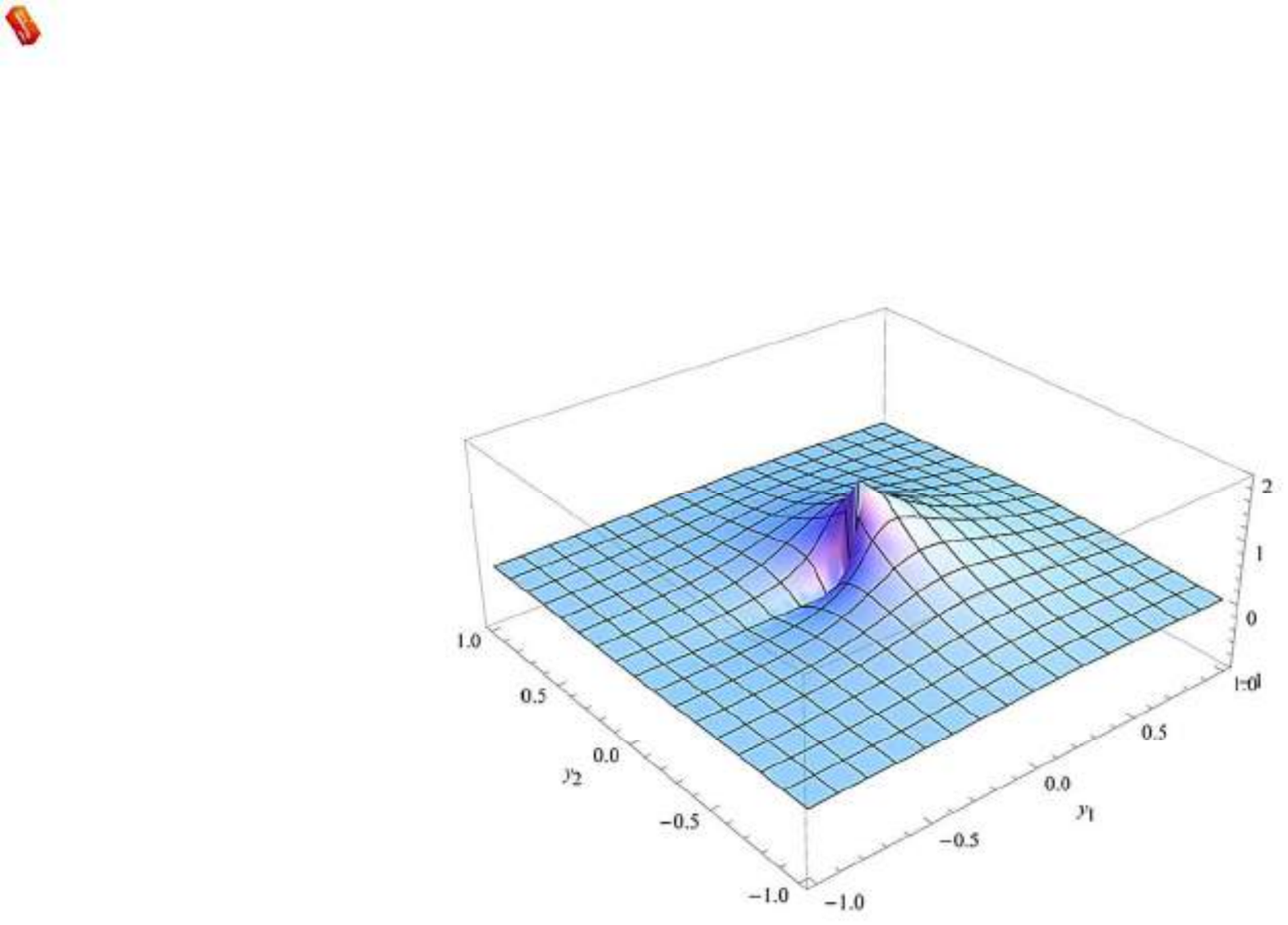}

\noindent{\it Fig.9} Wave function $\widetilde\Psi_{1,0}(\vec y)$ (see (\ref{ex-3})) of the first excited state for the Schr\"odinger equation (\ref{IIIi}) with potential (\ref{IIIk}) for Example III (case 2) with $a=1/8$ and $\omega_1=1,\, \omega_2=\sqrt{2}.$

\end{center}

\vspace{2pt}
\begin{center}
\includegraphics[width=0.8\textwidth]{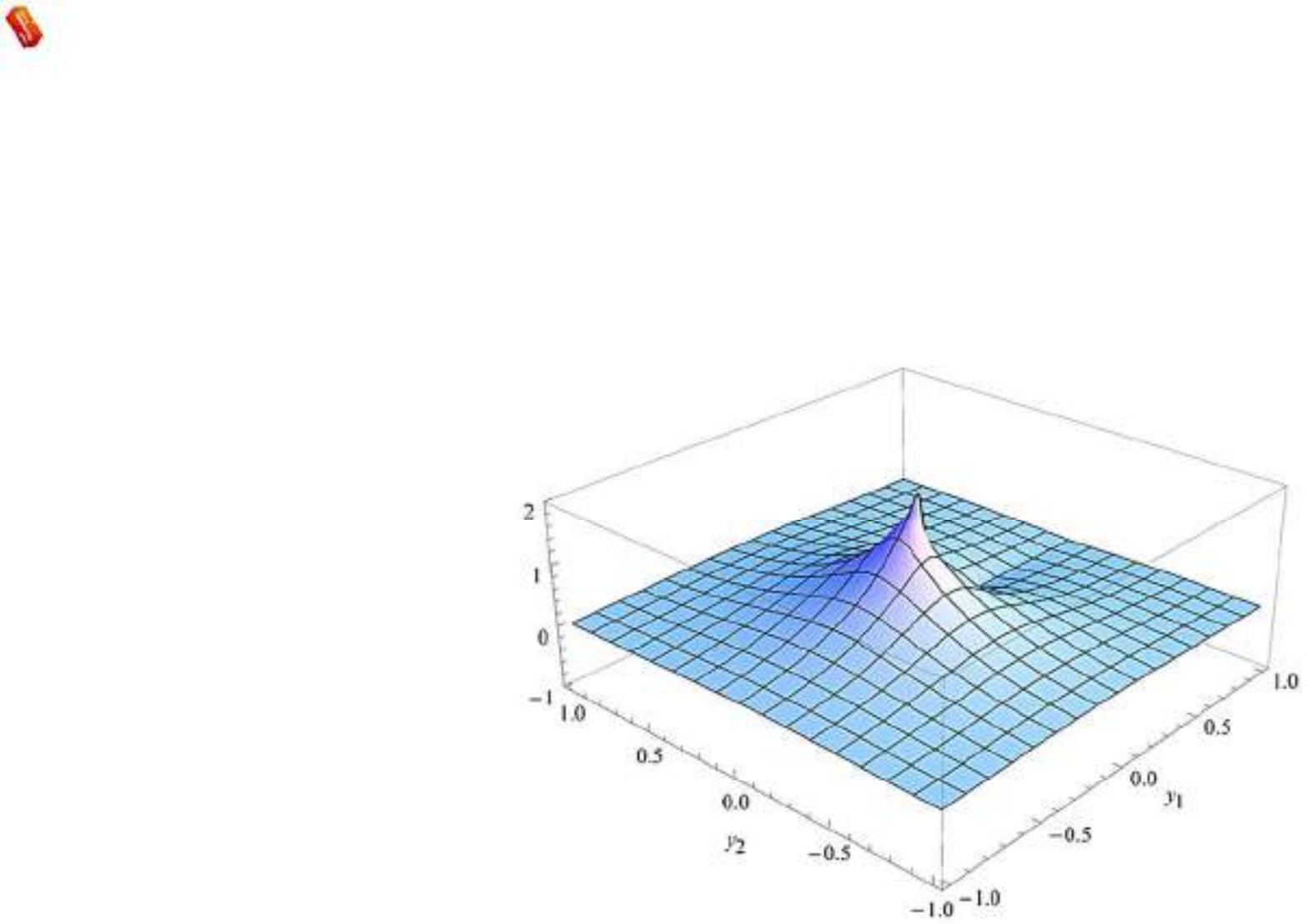}

\noindent{\it Fig.10} Wave function $\widetilde\Psi_{0,1}(\vec y)$ (see (\ref{ex-3})) of the second excited state for the Schr\"odinger equation (\ref{IIIi}) with potential (\ref{IIIk}) for Example III (case 2) with $a=1/8$ and $\omega_1=1,\, \omega_2=\sqrt{2}.$

\end{center}

\vspace{2pt}

{\bf IV.} Let us start by choosing anzats for the function $f(z)$:
\be
f(z)=\frac{1}{a\exp(-\lambda z) + b}=\frac{y_1-iy_2}{2}. \label{IVa}
\ee
Therefore,
\be
a\exp{(-\lambda x_1)}\biggl(\cos(\lambda x_2)-i\sin(\lambda x_2)\biggr)+b=\frac{2}{y_1-iy_2}=\frac{2y_1}{y_1^2+y_2^2}+\frac{2iy_2}{y_1^2+y_2^2},
\ee
providing the connection of $\vec x$ with polar coordinates $\rho,\,\varphi$ in the plane $\vec y:$
\ba
&&a^2\exp(-2\lambda x_1)=\frac{4}{\rho^2}-\frac{4b\cos\varphi}{\rho} +b^2;\nonumber\\
&&\cot(\lambda x_2)=\frac{b\rho}{2\sin\varphi}-\cot\varphi ;\quad \lambda >0. \label{IVb}
\ea
Function (\ref{IVa}) obeys the following relations:
\be
f'(z)=-\lambda f(z)\biggl(1-bf(z)\biggr);\quad f''(z)=-\lambda \biggl(1-2bf(z)\biggr)f'(z), \label{fff}
\ee
and according to (\ref{gM}) and (\ref{U}), the direct calculations give:
\ba
&&M(\vec y)=\frac{1}{4\lambda^2ff^{\star}\biggl(b^2ff^{\star}-b(f+f^{\star})+1\biggr)}=\frac{4}{\lambda^2\rho^2(b^2\rho^2-4b\rho\cos\varphi +4)};\label{oo}\\ &&U-V=-\frac{\lambda^2\biggl(1-2b(f+f^{\star})+4b^2ff^{\star}\biggr)}{4ff^{\star}}=-\frac{4(b^2\rho^2-2b\rho\cos\varphi +1)}{\rho^2(b^2\rho^2-4b\rho\cos\varphi +4)}. \label{IVc}
\ea
Taking into account Eqs.(\ref{IVb}), the initial potential $V(\vec x)$ can be taken as a combination of solvable one-dimensional potentials of Morse and trigonometrical Rosen-Morse models \cite{landau}, \cite{rosen}:
\be
V(\vec x)=V_1(x_1)+V_2(x_2)=C\biggl[\exp(-2\lambda x_1)
-2\exp(-\lambda x_1)\biggr]+
A\cot^2(\lambda x_2)+B\cot(\lambda x_2), \label{IVd}
\ee
with positive constants $A,B,C,\lambda .$ The resulting potential $U(\vec y)$ does not allow (for $b\neq 0$) the separation of variables, but nevertheless the solution of corresponding Schr\"odinger equation is expressed through solutions of one-dimensional problems with $V_1,\,V_2,$ which are known analytically \cite{landau}, \cite{rosen}.

Obtained solutions
may be used in different branches of Physics: from the traditional problems with PDM in Nuclear Physics and Condensed Matter Physics \cite{nucl} - \cite{-21-4} and till the modern problems of Cosmology where (\ref{ef}) can be interpreted \cite{tkachuk-1} as the Schr\"odinger equation in curved two-dimensional space. In particular, Example I might be useful for study of quantum particles in a medium with variable properties along one of the axis, and Example III might be related with the models of quantum dots for rotationally invariant effective mass of particles in external field.

\section{Conclusions}

For most of the problems with position dependent mass, the direct solution of the Schr\"odinger equation is impossible, especially for the spatial dimension $d>1.$ The situation becomes much better for the models obtained from the initial solvable constant mass problem by means of suitable point transformation. It was shown in the previous Sections, that the point transformations allow to obtain a wide class of solvable two-dimensional problems with different forms of dependence of mass function on coordinates.  In particular,
the initial problems amenable to separation of variables in Cartesian or some other coordinates create the class of corresponding solvable problems with PDM.

\section{Acknowledgments}

The work of M.V.I. was partially supported by the grant N 11.38.223.2015 of Saint Petersburg State University. We are grateful to the anonymous referee for useful remarks and advices. The plots in Figures 1-10 were built using the computer system MATHEMATICA.

\end{document}